\documentclass{article}
\usepackage{graphicx}
\title{\textbf{Perfect fluid quantum Universe in the presence of negative cosmological constant}}
\author{P. Pedram\thanks{Email: pouria.pedram@gmail.com}, M. Mirzaei, S. Jalalzadeh\thanks{Email: s-jalalzadeh@sbu.ac.ir}\,\, and S. S. Gousheh
\\ {\small Department of Physics, Shahid Beheshti University,
Evin, Tehran 19839, Iran}}
\begin{document}
\maketitle \baselineskip 24pt
\begin{abstract}
We present perfect fluid Friedmann-Robertson-Walker quantum cosmological models in the presence of
negative cosmological constant. In this work the Schutz's variational formalism is applied for
radiation, dust, cosmic string, and domain wall dominated Universes with positive, negative, and
zero constant spatial curvature. In this approach the notion of time can be recovered. These give
rise to Wheeler-DeWitt equations for the scale factor. We find their eigenvalues and eigenfunctions
by using Spectral Method. After that, we use the eigenfunctions in order to construct wave packets
for each case and evaluate the time-dependent expectation value of the scale factors, which are
found to oscillate between finite maximum and minimum values. Since the expectation values of the
scale factors never tends to the singular point, we have an initial indication that these models
may not have singularities at the quantum level.
\end{abstract}

\textit{Pacs}:{98.80.Qc, 04.40.Nr, 04.60.Ds} \maketitle

\section{Introduction}\label{sec1}
Quantum cosmological models are important subjects on the interface
of cosmology and gravitation. At first, B. DeWitt \cite{1} quantized
a Friedmann Universe filled with dust and later, closed isotropic
cosmological models with matter as a conformal and minimally coupled
scalar fields were quantized \cite{2,3}. Misner worked on the
quantization of anisotropic cosmological models \cite{4}, and
Barabanenkov quantized the Friedmann metric matched with the Kruskal
one \cite{5}. The quantization of a dust-like closed isotropic
cosmological model with a cosmological constant is also investigated
in Ref. \cite{6}.

In the quantum cosmology the Wheeler-DeWitt (WD) equation which
determines the wave function of the Universe, can be constructed
using ADM decomposition of the geometry \cite{7} in the Hamiltonian
formalism of general relativity. However, quantum cosmology has many
technical and conceptual problems. In fact, the WD equation of
quantum gravity is a functional differential equation defined in the
superspace which is the space of all possible three dimensional
spatial metrics, and no general solution is known in this
superspace. In quantum cosmology this problem is avoided by using
symmetry requirements to freeze out an infinite number of degrees of
freedom, leaving only a few for quantization process. This procedure
defines a minisuperspace, where exact solutions can often be found.
On the other hand, the general covariance will be lost upon applying
the ADM decomposition, and in most cases the notion of time
disappears at the quantum level \cite{8}. Even, if all these
problems are solved, the interpretation of the central object, i.e.
the wave function of the Universe, remains unanswered.

The many-worlds interpretations \cite{9} of quantum mechanics is one
of the most popular interpretation schemes for the wave function of
the Universe. This interpretation differs noticeably from the
Copenhagen interpretation of quantum mechanics since the conception
of probability is abandoned in some sense. In fact, all
possibilities are participated to create new Universes with
different possible eigenvalues obtained by measurements. The
evolution of observables such as scale factor is found by evaluating
the expectation values.  In this case, like in the Copenhagen
interpretation, the structure of Hilbert space and self-adjoint
operators are still unchanged.

The presence of the matter in quantum cosmology needs further
consideration and can be described by fundamental fields, as done in
Ref. \cite{10}. Using WKB approximation one can predict the behavior
of the quantum Universe which leads to determination of the
trajectories in phase space. However, even in the minisuperspace,
general exact solutions are hard to find, the Hilbert space
structure is ambiguous and it is difficult to recover the conception
of a semiclassical time \cite{8,10}.

Here, we consider matter as a perfect fluid. This description is
basically semiclassical, but it introduces a variable, which can be
identified with time and connected with the matter degrees of
freedom, leading to a well-defined Hilbert space structure.
Moreover, this allow us to treat the barotropic equation of state $p
= \alpha\rho$ with arbitrary $\alpha$.

It is very convenient to construct a quantum perfect fluid model.
Schutz's formalism \cite{11,12} gives dynamics to the fluid degrees
of freedom in interaction with the gravitational field. Using a
proper canonical transformations, at least one conjugate momentum
operator associated with matter appears linearly in the action
integral. Therefore, a Schr\"odinger-like equation can be obtained
where the matter plays the role of time. Moreover, recently, some
applications of the Schutz's formalism have been discussed in the
framework of the perfect fluid Stephani Universe
\cite{pedramPLB,pedramCQG2} and Friedmann-Robertson-Walker (FRW)
Universe in the presence of Chaplygin gas \cite{pedramIJTP,pedramPLB2}.

Until now, quantum perfect fluid models with common equations of
state have been constructed in the absence of cosmological constants
\cite{13,14,15,16,17}. We can study the behavior of the scale factor
using the many-worlds and the de Broglie-Bohm interpretations of
quantum mechanics.

Recently, the quantization of FRW radiation dominated Universe in
the presence of a negative cosmological constant is discussed by
Monerat \textit{et al} \cite{18}. However, as mentioned in Ref.
\cite{19}, their results are inaccurate and their relative errors
range between $10^{-3}$ for the ground state of $k=1$  case, and $1$
for the ground state of $k=-1$, which make their work unreliable.
Here, we generalize the previous investigations by studying quantum
perfect fluid models for barotropic equation of state $p =
\alpha\rho$, where $\alpha=\{1/3,0,-1/3,-2/3\}$ correspond to
radiation, dust, cosmic string, and domain wall dominated Universes,
respectively. Using the many-worlds framework, the behavior of the
scale factor is determined, although the results are independent of
the interpretation scheme employed. The large time average of the
expectation value of the scale factor is similar to the classical
case. Moreover, the model predicts an accelerated expansion today if
$- 1/3 > \alpha > - 1$.

It is important to mention that although recent observations point
toward a positive cosmological constant, it is still possible that
at the very early Universe the cosmological constant be negative.
Moreover, we think it is important to understand more about such
models which represent bound Universes.

This paper is organized as follows.  We quantize three
Friedmann-Robertson-Walker perfect fluid models in the presence of a
negative cosmological constant, using the formalism of quantum
cosmology. In Sec.~\ref{sec2}, the quantum cosmological model with a
perfect fluid as the matter content is constructed in Schutz's
formalism \cite{11,12}, and the WD equation in  minisuperspace is
Found to quantize the model. The wave-function depends on the scale
factor $a$ and on the canonical variable associated to the fluid
which plays the role of time $T$, in the Schutz's variational
formalism. We separate the wave-function in two parts, one depending
solely on the scale factor and the other depending only on the time.
The solution in the time sector of the WD equation is trivial,
leading to imaginary exponentials of the type $e^{iEt}$, where $E$
is the system energy and $t =T$. In Sec.~\ref{sec3}, we outline the
Spectral Method \cite{boyd,SMP,25}, and use it to find the
eigenvalues and eigenfunctions of corresponding WD equations. In
Sec.~\ref{sec4}, we construct wave packets from the eigenfunctions,
for radiation, dust, cosmic string and domain wall dominated
Universes respectively, and compute the time-dependent expectation
values of the scale factors for $k=1,0,-1$. In Sec.~\ref{sec5}, we
state our conclusions.

\section{Model}\label{sec2}
Let us start from the Einstein-Hilbert action plus a perfect fluid
in the formalism developed by Schutz. For this, we write down the
action for gravity plus perfect fluid as
\begin{eqnarray}
\label{action} S &=& \frac{1}{2}\int_Md^4x\sqrt{-g}\, (R-2\Lambda) +
2\int_{\partial M}d^3x\sqrt{h}\, h_{ab}\, K^{ab} +
\int_Md^4x\sqrt{-g}\, p,
\end{eqnarray}
here, $K^{ab}$ is the extrinsic curvature, $\Lambda$ is the
cosmological constant, and $h_{ab}$ is the induced metric over the
three-dimensional spatial hypersurface, which is the boundary
$\partial M$ of the four dimensional manifold $M$. We choose units
such that the factor $8\pi G$ becomes equal to one. The first two
terms were first obtained in \cite{7} and the last term of
(\ref{action}) represents the matter contribution to the total
action, $p$ being the pressure which obeys the barotropic equation
of state $p = \alpha\rho$. In Schutz's formalism \cite{11,12} the
fluid's four-velocity can be expressed in terms of five potentials
$\epsilon$, $\zeta$, $\beta$, $\theta$ and $S$
\begin{equation}
u_\nu = \frac{1}{\mu}(\epsilon_{,\nu} + \zeta\beta_{,\nu} + \theta
S_{,\nu})
\end{equation}
where $\mu$ is the specific enthalpy. $S$ is the specific entropy,
and the potentials $\zeta$ and $\beta$ are connected with rotation
which are absent of models in the FRW type. The variables $\epsilon$
and $\theta$ have no clear physical meaning. The four-velocity also
satisfies the normalization condition
\begin{equation}
u^\nu u_\nu = -1.
\end{equation}
The FRW metric
\begin{equation}
ds^2 = - N^2(t)dt^2 + a^2(t)g_{ij}dx^idx^j,
\end{equation}
can be inserted in the action (\ref{action}), where $N(t)$ is the
lapse function and $g_{ij}$ is the metric on the constant-curvature
spatial section. After some thermodynamical considerations and using
the constraints for the fluid, and dropping the surface terms, the
final reduced action takes the form \cite{14}.
\begin{equation}
S = \int dt\biggr[-3\frac{\dot a^2a}{N} -\Lambda N a^3+ 3kNa +
N^{-1/\alpha} a^3\frac{\alpha}{(\alpha + 1)^{1/\alpha +
1}}(\dot\epsilon + \theta\dot S)^{1/\alpha + 1}\exp\biggr(-
\frac{S}{\alpha}\biggl) \biggl].
\end{equation}
The reduced action may be further simplified using canonical methods
\cite{14}, resulting in the super-Hamiltonian
\begin{equation}
{\cal H} = - \frac{p_a^2}{12a}+\Lambda a^3 -3ka +\frac{
 p_\epsilon^{\alpha + 1}e^S}{a^{3\alpha}},
\end{equation}
where $p_a= -6{\dot aa}/{N}$ and $p_\epsilon = -\rho_0 u^0 Na^3$,
$\rho_0$ being the rest mass density of the fluid. The following
additional canonical transformations, which generalizes the one
used in \cite{14},
\begin{eqnarray}
T &=& -p_Se^{-S}p_\epsilon^{-(\alpha + 1)}, \quad  \quad p_T
=p_\epsilon^{\alpha + 1}e^S \quad , \quad\nonumber\\
\bar\epsilon &=& \epsilon - (\alpha + 1)\frac{p_S}{p_\epsilon},
\quad \quad \quad \bar p_\epsilon = p_\epsilon,
\end{eqnarray}
simplifies the super-Hamiltonian to,
\begin{equation}
{\cal H} = - \frac{p_a^2}{12a} +\Lambda a^3- 3ka +
\frac{p_T}{a^{3\alpha}} \,\, ,\label{EqHamiltonian}
\end{equation}
where the momentum $p_T$ is the only remaining canonical variable
associated with matter and appears linearly in the
super-Hamiltonian. The parameter $k$ defines the curvature of the
spatial section, taking the values $0, 1, - 1$ for a flat, close or
open Universes, respectively.

The classical dynamics is governed by the Hamilton equations,
derived from Eq. (\ref{EqHamiltonian}) and Poisson brackets as
\begin{equation}
\left\{
\begin{array}{llllll}
\dot{a} =&\{a,N{\cal H}\}=-\frac{\displaystyle Np_{a}}{\displaystyle 6a}\, ,\\
 & \\
\dot{p_{a}} =&\{p_{a},N{\cal H}\}=- \frac{N}{12a^2}p_a^2+3Nk  \\
& \\
&-3N\Lambda a^2+N3\alpha a^{-3\alpha-1}p_T \, ,\\
& \\
\dot{T} =&\{T,N{\cal H}\}=Na^{-3\alpha}\, ,\\
 & \\
\dot{p_{T}} =&\{p_{T},N{\cal H}\}=0\, .\\
& \\
\end{array}
\right. \label{4}
\end{equation}
We also have the constraint equation ${\cal H} = 0$. Choosing the
gauge $N=a(t)$, we have the following solutions for the system
\begin{eqnarray}
\ddot{a}&=&-ka+\frac{2}{3}\Lambda
a^3+\frac{1-3\alpha}{6}a^{-3\alpha}p_T,\\
0&=&-3\dot a^2+\Lambda a^4-3k a^2+a^{1-3\alpha}p_T.
\end{eqnarray}
The classical equation of motion for the scale factor in absent of
the cosmological constant is solved in a unified form for any
$\alpha\in[0,1]\,$ in terms of hypergeometric functions in Ref.
\cite{assad}. Moreover, In the radiation dominated Universe
($\alpha=1/3$) with a negative cosmological constants, the classical
solutions have been obtained using Jacobi's elliptic sine functions
\cite{18}. the WD equation in minisuperspace can be obtained by
imposing the standard quantization conditions on the canonical
momenta and ($p_a \rightarrow -i\frac{\displaystyle
\partial}{\displaystyle\partial a}$,  $p_T \rightarrow
-i\frac{\displaystyle\partial}{\displaystyle\partial T}$) demanding
that the super-Hamiltonian operator annihilate the wave function
($\hbar =1$)
\begin{equation}
\label{sle} \frac{\partial^2\Psi}{\partial a^2}+12\Lambda a^4\Psi -
36ka^2\Psi - i12a^{1 - 3\alpha}\frac{\partial\Psi}{\partial t} = 0.
\end{equation}
where $t=T$ corresponds to the time coordinate. Equation (\ref{sle})
takes the form of a Schr\"odinger equation $i\partial\Psi/\partial t
= {\hat H} \Psi$. Demanding that the Hamiltonian operator ${\hat H}$
to be self-adjoint, the inner product of any two wave functions
$\Phi$ and $\Psi$ must take the form \cite{nivaldo,15}
\begin{equation}
(\Phi,\Psi) = \int_0^\infty a^{1 - 3\alpha}\Phi^*\Psi da,
\end{equation}
On the other hand, the wave functions should satisfy the following
boundary conditions
\begin{equation}
\label{boundary} \Psi(0,t) = 0 \quad \mbox{or} \quad
\frac{\partial\Psi (a,t)}{\partial a}\bigg\vert_{a = 0} = 0.
\end{equation}
The WD equation (\ref{sle}) can  be solved by separation of
variables as follows
\begin{equation}
\psi(a,t) = e^{iEt}\psi(a), \label{11}
\end{equation}
where the $a$ dependent part of the wave function ($\psi(a)$)
satisfies
\begin{equation}
\label{sle2} -\psi''(a) +( 36 ka^2-12\Lambda a^4)\psi(a) =12Ea^{1 -
3\alpha}\psi(a).
\end{equation}
Since the energy term grows faster than the potential for $\alpha <
- 1$, this equation  has a discrete spectra ($E_n$) with associated
bound state eigenfunctions ($\psi_n(x)$) only for $\alpha > -1$.

We construct a general solution to the WD equation (\ref{sle}) by
taking linear combinations of the $\psi_{n} (a, t)$'s,
\begin{equation}
\Psi (a,t) = \sum_{n=0}^{m} C_{n}(E_n)\psi_n(a) e^{iE_{n}t},
\label{wavepacket}
\end{equation}
where the coefficients $C_n(E_n)$ will be fixed later. Form pure
mathematical point of view, by allowing negative values of $a$, the
Parity operator can be defined. If the WD equation (Eq. \ref{sle2})
is covariant under the Parity operator, its eigenfunctions can be
separated into even and odd ones. The even or odd wave packets
constructed from appropriate linear combinations of the eigenstates,
have the important property that they will not change their parity
in the course of their time evolution. Therefore, if we choose the
initial wave packets to be odd or even, that is they satisfy either
the first or  the second condition stated in Eq. \ref{boundary}
respectively, they will satisfy them for all times. We can compute
the expected value for the scale factor $a$ for any wave function,
using the {\it many worlds interpretation} of quantum mechanics.
This means, we may write the expected value for the scale factor $a$
as \cite{22}
\begin{equation}
\left<a\right>_t = \frac{\int_{0}^{\infty}a^{2-3\alpha}\,|\Psi
(a,t)|^2 da} {\int_{0}^{\infty}a^{1-3\alpha}\,|\Psi(a,t)|^2 da}.
\label{meanvalue}
\end{equation}

Before solving the WD equation (\ref{sle2}) via Spectral Method, it
is worthy to state a brief overview of the Chhajlany and Malnev
method and Variational Sinc Collocation Method (VSCM), which have
been recently used to solve the WD equation (\ref{sle2}), for
radiation epoch ($\alpha=1/3$) in Refs. \cite{18,19}, respectively.

In Chhajlany and Malnev method \cite{Chhajlany1,Chhajlany2}, one
adds an extra term to the original anharmonic oscillator potential
to find a subset of normalizable solutions of the modified
Hamiltonian. In the case of equation (\ref{sle2}) this extra term is
$c\,a^6$ where $c$ is constant. Now, the solution can be written as
a polynomial where the larger the degree of the polynomial, the
smaller the constant, $c$ is. In fact, by increasing the order of
polynomial, the energy eigenvalues predicted by this method approach
monotonically to the energy eigenvalues of the original Hamiltonian.

On the other hand, to obtain highly accurate numerical results, both
for the energy eigenvalues and eigenfunctions, one can use
Variational Sinc Collocation Method (VSCM) \cite{Amore}. It is shown
that the errors decay exponentially with the number of elements
(sinc functions) used for discretization of the Hamiltonian.
Diagonalization of the resulting matrix, by specification of the
otherwise arbitrary grid spacing $h$ (spacing between two contiguous
sinc functions), yields energy eigenvalues and eigenfunctions. As
shown by Amore {\it et al}, for a specified number of sinc
functions, there exists an optimal value of $h$ which yields the
minimum errors \cite{Amore}. this optimal value can be found using
the Principle of Minimal Sensitivity (PMS) \cite{Stevenson} to the
trace of the Hamiltonian matrix.

As indicated by Lemos {\it et al}, the need for a modified potential
instead of the original one in the Chhajlany and Malnev method,
gives rise to significant errors, particularly for $k=-1$
\cite{Reply}. In fact, VSCM is more uniformly accurate and converges
more rapidly than the Chhajlany and Malnev method.

\section{The Spectral Method}\label{sec3}
In this section we introduce Spectral Method (SM) \cite{boyd,SMP} as
a tool for solving differential equation. We have recently used the
Spectral Method for constructing the appropriate wave packets which
are solutions to a WD equation \cite{25}. This method is simple,
fast, accurate and stable.

Let us consider the general time-independent WD equation (EQ.
(\ref{sle2})),
\begin{equation}\label{RSM1}
-\frac{d^2\psi(x)}{dx^2}+\hat{f}[x]\psi(x)=E\,\hat g[x]\,\,
\psi(x),
\end{equation}
where $\hat{f}$ and $\hat g$ are arbitrary, but with derivative
operators less than two. For the usual eigenvalue problem $\hat
g=1$, which includes the time-independent Schr\"{o}dinger equation.
The method SM can be easily extended to solve the general case which
$\hat g$ is a operator in the $x$ space. This generalize problem can
be named a generalized eigenvalue problem. Throughout this paper, we
only examine the bound states of this problem, i.e. the states which
are the square integrable. The configuration space for most physical
problems are defined by $-\infty<x<\infty$. Since the bound states
fall off sufficiently fast for large $|x|$, a finite region
suffices, and the proper choice for this region, say $-L/2<x<L/2$.
The use of a finite domain is also necessary since we need to choose
a finite subspace of a countably infinite basis. We find it
convenient to shift the domain to $0<x<L$. In particular, we need to
shift the potential energy functions also. This means that we can
expand the solution as,
\begin{eqnarray} \psi(x)=\sum_{n=1}^{\infty} A_{n}\sqrt{\frac{2}{L}}
\,\,\, \sin\left(\frac{n \pi x}{L}\right).
\label{eqpsitrigonometric}
\end{eqnarray}
We can also make the following expansions,
\begin{eqnarray}
\hat f\,\, \psi(x)&=&\sum_{n} B_{n}
\,\,\,\sqrt{\frac{2}{L}}\sin\left(\frac{n \pi
x}{L}\right),\label{eqV1}\\
\hat g\,\, \psi(x)&=&\sum_{n} B'_{n}
\,\,\,\sqrt{\frac{2}{L}}\sin\left(\frac{n \pi
x}{L}\right),\label{eqV2}
\end{eqnarray}
where $B_{n}$ $B'_{n}$ are coefficients that can be determined
once $\hat f$ and $\hat g$ are specified. By substituting Eqs.\
(\ref{eqpsitrigonometric},\ref{eqV1},\ref{eqV2}) into Eq.\
(\ref{RSM1}) and using the differential equation of the Fourier
basis we obtain,
\begin{eqnarray}
\sum_{n}\left[\left(\frac{n \pi}{L}\right)^2
A_{n}+B_{n}\right]\sin\left(\frac{n \pi
x}{L}\right)=E\sum_{n}B'_n\sin\left(\frac{n \pi
x}{L}\right).\label{eqAB1}
\end{eqnarray}
Because of the linear independence of $\sin\left(\frac{n \pi
x}{L}\right)$, every term in the summation must satisfy,
\begin{eqnarray}
\left(\frac{n \pi}{L}\right)^2 A_{n}+B_{n}=E\,
B'_{n}.\label{eqAB2}
\end{eqnarray}
It only remains to determine the matrices $B$ and $B'$. Using
Eqs.\ (\ref{eqV1},\ref{eqV2}) and Eq.\ (\ref{eqpsitrigonometric})
we have,
\begin{eqnarray}
\sum_{n} B_{n}\sin\left(\frac{n \pi x}{L}\right)=\sum_{n} A_{n}
\hat f\,\, \sin\left(\frac{n \pi x}{L}\right),\\
\sum_{n} B'_{n}\sin\left(\frac{n \pi x}{L}\right)=\sum_{n} A_{n}
\hat g\,\, \sin\left(\frac{n \pi x}{L}\right),
\end{eqnarray}
By multiplying both sides of the above equations by
$\sin\left(\frac{n \pi x}{L}\right)$ and integrating over the
$x$-space and using the orthonormality condition of the basis
functions, one finds,
\begin{eqnarray}
B_{n}= \sum_{m} C_{m,n}\,\, A_{m},\\
B'_{n}= \sum_{m} C'_{m,n}\,\, A_{m},
\end{eqnarray}
where,
\begin{eqnarray}
C_{m,n}=\frac{2}{L}\int_{0}^{L}\sin\left(\frac{m \pi
x}{L}\right)\hat f\,\,\sin\left(\frac{n \pi x}{L}\right) dx,\\
C'_{m,n}=\frac{2}{L}\int_{0}^{L}\sin\left(\frac{m \pi
x}{L}\right)\hat g\,\,\sin\left(\frac{n \pi x}{L}\right) dx,
\end{eqnarray}
Therefore we can rewrite Eq.\ (\ref{eqAB2}) as,
\begin{eqnarray}
\left(\frac{n \pi}{L}\right)^2 A_{n}+ \sum_{m} C_{m,n}\,\, A_{m}=E
\sum_{m} C'_{m,n}\,\, A_{m}.\label{eqAC}
\end{eqnarray}
It is obvious that the presence of the operators $\hat f$ and
$\hat g$ in Eq.\ (\ref{RSM1}), leads to nonzero coefficients
$C_{m,n}$ and $C'_{m,n}$ in Eq.\ (\ref{eqAC}), which in principle
could couple all of the vector elements of $A$. It is easy to see
that the more basis functions we include, the closer our solution
will be to the exact one. We select a finite subset of the basis
functions \textit{i.e.} the first $N$ ones, by letting the index
$m$ run from 1 to $N$ in the summations.  Equation (\ref{eqAC})
can be written as,
\begin{eqnarray}
D\, A=E \,D'\, A, \label{eqmatrix}
\end{eqnarray}
or,
\begin{eqnarray}
D'^{-1}D\, A=E \, A, \label{eqmatrix2}
\end{eqnarray}
where $D$ and $D'$ are square matrices with $N \times N$ elements.
Their elements can be obtained from Eq.~(\ref{eqAC}). The solution
to this matrix equation simultaneously yields $N$ sought after
eigenstates and eigenvalues. It is important to note that the
optimized value of $L$ crucially depends on the number of basis
functions $N$ ($L(N)$), which results in the maximum accuracy and
the stability of the solutions (for a comprehensive study about the
optimization procedure see \cite{SMP}).

\section{Results}\label{sec4}
In this section we will solve the equation (\ref{sle2}) using SM. By
choosing $N=100$ basis functions, and we report our results with 10
significant digits. Note that, although, we are free to choose other
values of $\Lambda$, but the accuracy of results for small
$|\Lambda|$ reduces in comparison with large values of $|\Lambda|$
for a given number of basis $N$, particularly for $k=-1$. This means
that we need to increase the number of basis $N$ to obtain the same
accuracy which increases the computations. With regard to these
considerations, the results are robust under changes of Lambda.

\subsection{Radiation ($\alpha=1/3$)}
\begin{figure}
\centering
  % Requires \usepackage{graphicx}
  \includegraphics[width=7cm]{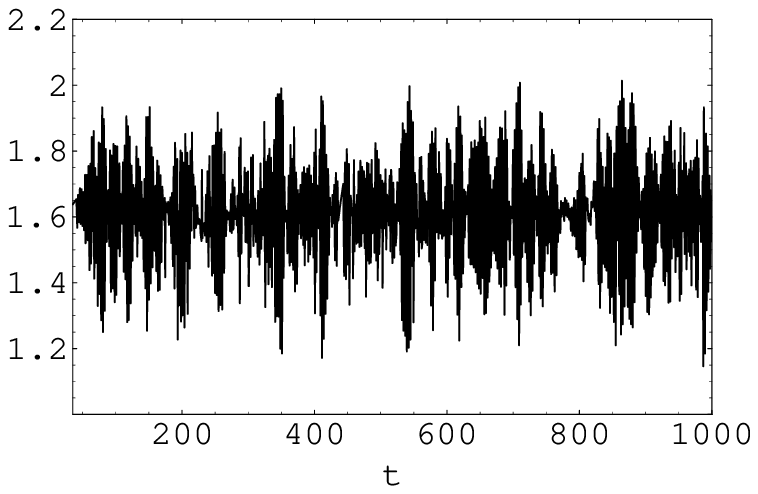}
  \caption{Behavior of the expectation value of the scalar factor for $\Lambda=-0.1$,
  $k=1$, and $C_n=1,0$ for odd and even $n$,
respectively, in radiation regime.}\label{figRad1}
\end{figure}

\begin{figure}
\centering
  % Requires \usepackage{graphicx}
  \includegraphics[width=7cm]{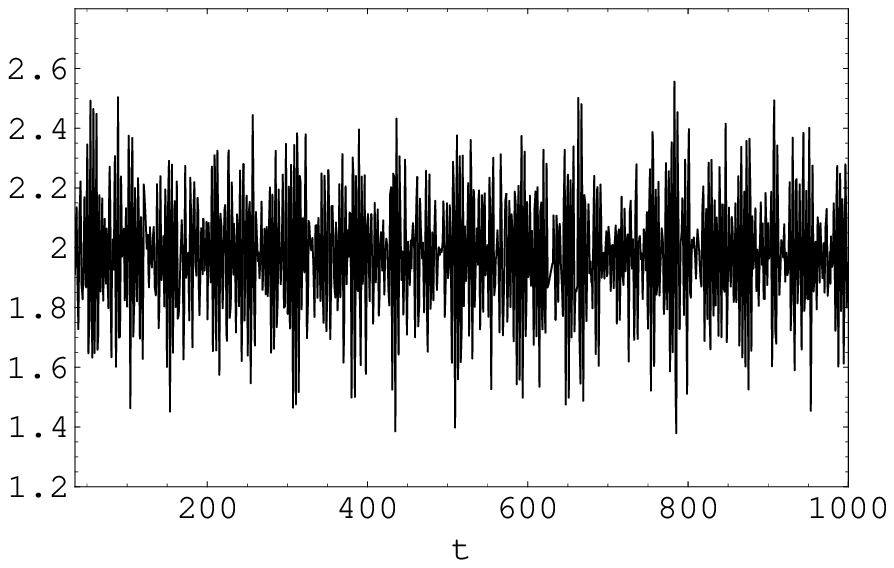}\\
  \caption{Behavior of the expectation value of the scalar factor for $\Lambda=-0.1$,
  $k=0$, and $C_n=1,0$ for odd and even $n$,
respectively, in radiation regime.}\label{figRad0}
\end{figure}

\begin{figure}
\centering
  % Requires \usepackage{graphicx}
  \includegraphics[width=7cm]{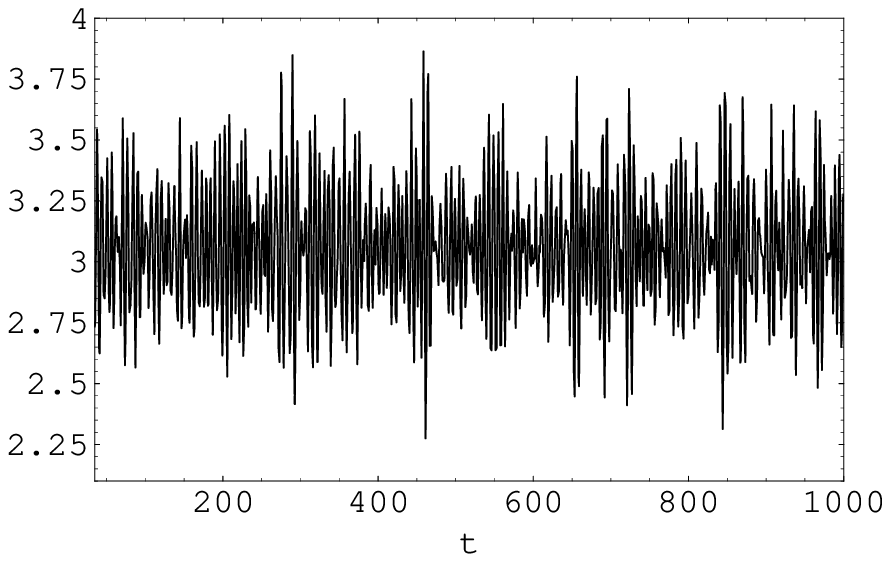}\\
  \caption{Behavior of the expectation value of the scalar factor for $\Lambda=-0.1$,
  $k=-1$, and $C_n=1,0$ for odd and even $n$,
respectively, in radiation regime.}\label{figRad-1}
\end{figure}

\begin{figure}
\centering
  % Requires \usepackage{graphicx}
  \includegraphics[width=7cm]{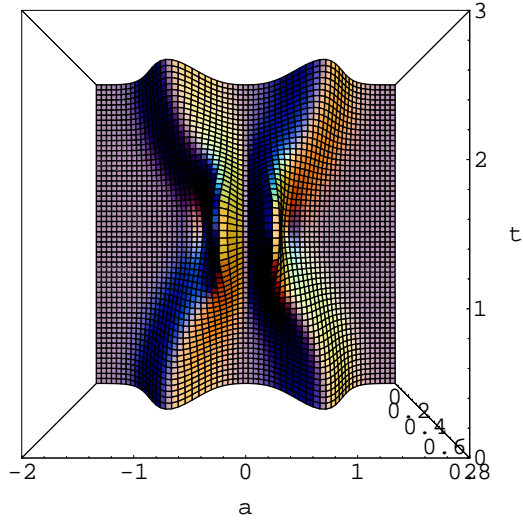}
  \caption{3D plot of the square of the wave packet for $k=1$ in radiation regime with the coefficients of odd coherent state of the quantum simple
harmonic oscillator.}\label{figplot}
\end{figure}

In the radiation dominated Universe time-independent WD equation has
the following form,
\begin{equation}
\label{Radiation} -\frac{d^2\psi(a)}{da^2} +( 36 ka^2-12\Lambda
a^4)\psi(a) =12E\psi(a),
\end{equation}
In this form it is obvious that the system is absolutely stable for $\Lambda<0$. Note that equation
is covariant under the Parity operator. For ease of comparison of our results with those of Refs.
\cite{18,19}, we select the first condition of the equation (\ref{boundary}) and choose the
coefficients $C_n\,$s in equation (\ref{wavepacket}) to be 1 and zero for the odd and even
eigenfunctions, respectively. We can find the energy eigenvalues and eigenvectors of this equation
with ease using SM where $\hat{f}=36 kx^2-12\Lambda x^4 $ and $\hat{g}=12 $ in comparison with Eq.
(\ref{RSM1}). Table \ref{TabRadiation} shows the first 26 odd eigenvalues for $k=1,0,-1$ respectively. Figures
(\ref{figRad1},\ref{figRad0},\ref{figRad-1}) show the resulting expectation values of the scale
factor $a$, versus $t$ for the various values of $k$. As can be seen from the table, the results
are as same as those reported in Ref. \cite{19}. To show the arbitrariness in choosing initial odd
wave packet, we can use the coefficients of odd coherent state of the quantum simple harmonic
oscillator. Figure \ref{figplot} shows the 3D plot of resulting wave packet for $k=1$ case.

\begin{table}

 \centering
\begin{tabular}{|c|c|c|c|}
  % after \\: \hline or \cline{col1-col2} \cline{col3-col4} ...
  \hline
   & $k=1$ & $k=0$ & $k=-1$ \\\hline
$E_{1}$&1.510262538\ &\
0.3364795921&-21.79569604\\

 \hline $E_{2}$&3.550647291\
&\ 1.031199050&-20.39848969\\

 \hline
$E_{3}$&5.621893706\ &\
1.880761581&-19.01885126\\

 \hline $E_{4}$&7.722777814\
&\ 2.842487493&-17.65745105\\

 \hline
$E_{5}$&9.852194220\ &\
3.894746211&-16.31503013\\

 \hline $E_{6}$&12.00913857\
&\ 5.024091561&-14.99241327\\

 \hline
$E_{7}$&14.19269339\ &\
6.221192430&-13.69052532\\

 \hline $E_{8}$&16.40201655\
&\ 7.479120856&-12.41041248\\

\hline $E_{9}$&18.63633173\ &\
8.792490724&-11.15327022\\

 \hline $E_{10}$&20.89492047\
&\ 10.15697184&-9.920481041\\

 \hline
$E_{11}$&23.17711552\ &\
11.56899282&-8.713666752\\

\hline $E_{12}$&25.48229516\
&\ 13.02554815&-7.534763378\\

\hline $E_{13}$&27.80987836\ &\
14.52406697&-6.386132854\\

 \hline $E_{14}$&30.15932069\
&\ 16.06232038&-5.270738293\\

 \hline
$E_{15}$&32.53011067\ &\
17.63835396&-4.192437789\\

\hline $E_{16}$&34.92176669\
&\ 19.25043737&-3.156519423\\

 \hline
$E_{17}$&37.33383429\ &\
20.89702582&-2.170719225\\

\hline $E_{18}$&39.76588373\
&\ 22.57673016&-1.246288323\\

\hline $E_{19}$&42.21750793\ &\
24.28829319&-0.3899963301\\

\hline $E_{20}$&44.68832056\
&\ 26.03057067&0.4337198672\\

\hline $E_{21}$&47.17795441\ &\
27.80251593&1.300741394\\

 \hline $E_{22}$&49.68605987\
&\ 29.60316715&2.243855391\\

 \hline
$E_{23}$&52.21230364\ &\
31.43163692&3.256001774\\

\hline $E_{24}$&54.75636750\
&\ 33.28710339&4.326509414\\

 \hline
$E_{25}$&57.31794728\ &\
35.16880284&5.448058248\\

\hline $E_{26}$&59.89675181\
&\ 37.07602341&6.615611361\\

\hline
\end{tabular}
\caption{The lowest calculated energy levels for the cases $k=0$,
$k=1$, and $k=-1$ in radiation dominated Universe (in all cases,
$\Lambda=-0.1$).}\label{TabRadiation}
\end{table}

\subsection{Dust ($\alpha=0$)}
\begin{figure}
\centering
  % Requires \usepackage{graphicx}
  \includegraphics[width=7cm]{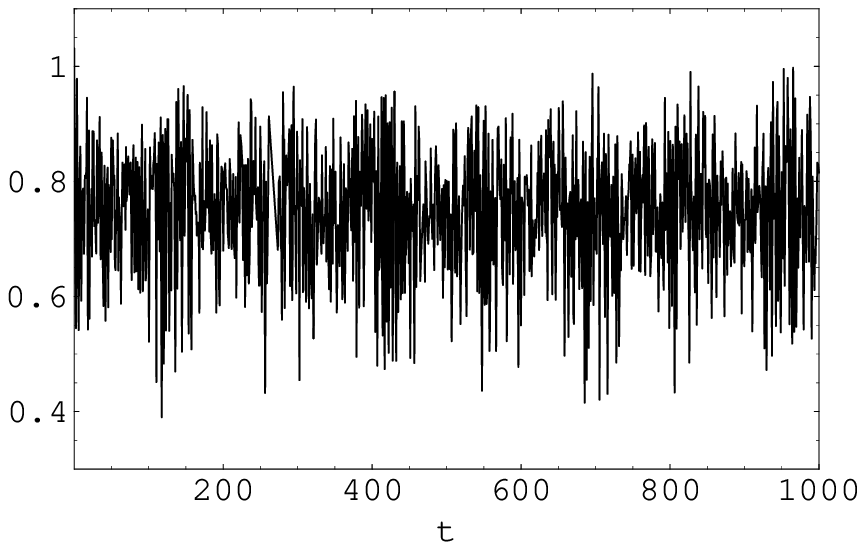}
  \caption{Behavior of the expectation value of the scalar factor for $\Lambda=-15$,
  $k=1$, and $\Psi(a,0) =
\exp(-8(a-1)^2)$ in dust regime.}\label{figDust1}
\end{figure}

\begin{figure}
\centering
  % Requires \usepackage{graphicx}
  \includegraphics[width=7cm]{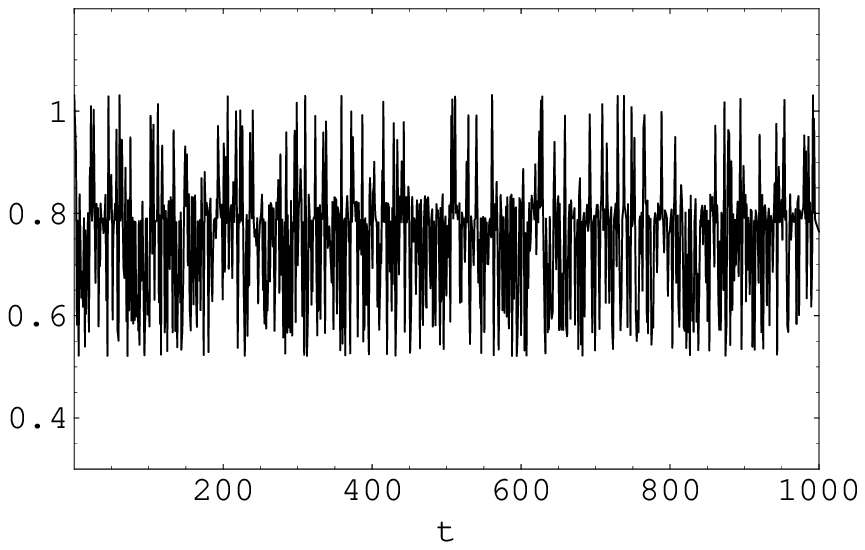}\\
  \caption{Behavior of the expectation value of the scalar factor for $\Lambda=-15$,
  $k=0$, and $\Psi(a,0) =
\exp(-8(a-1)^2)$ in dust regime.}\label{figDust0}
\end{figure}

\begin{figure}
\centering
  % Requires \usepackage{graphicx}
  \includegraphics[width=7cm]{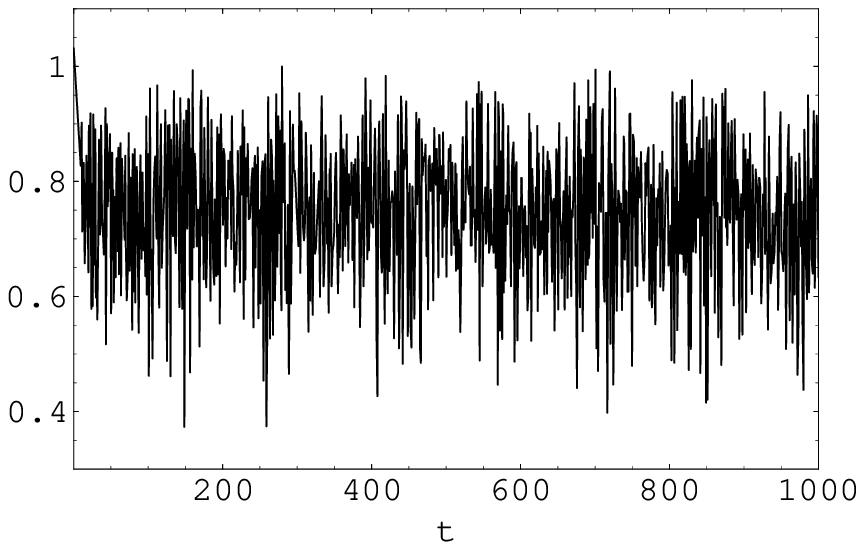}\\
  \caption{Behavior of the expectation value of the scalar factor for $\Lambda=-15$,
  $k=-1$, and $\Psi(a,0) =
\exp(-8(a-1)^2)$ in dust regime.}\label{figDust-1}
\end{figure}

In dust dominated Universe time-independent WD equation has the
following form,
\begin{equation}
\label{Dust} -\frac{d^2\psi(a)}{da^2} +( 36 ka^2-12\Lambda
a^4)\psi(a) =12Ea\psi(a),
\end{equation}
We can find the energy eigenvalues and eigenvectors of this equation with ease using SM where
$\hat{f}=36 kx^2-12\Lambda x^4 $ and $\hat{g}=12x $ in notation displayed in Eq. (\ref{RSM1}).
Table \ref{TabDust} shows the first 20 positive eigenvalues for $k=1,0,-1$ respectively. Note that, for any
positive eigenvalues ($E_n^+$), there is an negative counterpart ($E_n^-$) which $E_n^-=-E_n^+$.
The above equation, is not invariant under the Parity operator. Therefore, its eigenfunctions can
not in general satisfy either of the conditions of equation (\ref{boundary}. However, we can
construct wave packets, from linear combinations of the eigenfunctions, which vanishes at $a=0$ and
$t=0$. Then we need to check that the constraints (Eq. \ref{boundary} remain valid for all $t$ for
our choice of initial condition. For example we can choose the coefficients $C_n\, $s so as to
construct a gaussian initial wave packet ($\Psi(a,0)$) which is centered \textit{e.g.} at $a=1$.
Figures (\ref{figDust1},\ref{figDust0},\ref{figDust-1}) show the resulting expectation values of
the scale factor $a$, versus $t$ for the various values of $k$. As can seen from the figures these
wave packets always satisfy the first boundary condition (Eq. \ref{boundary}).

In the case $\Lambda=0$, the time-independent WD equation
(\ref{Dust}) reduces to
\begin{equation}
-\frac{d^2\psi(a)}{da^2} + 36 ka^2\psi(a) =12Ea\psi(a),
\label{Dust1}
\end{equation}
In terms of the new variable $x=6a - E$ we find
\begin{equation}
-\frac{d^2\psi(x)}{dx^2} +
\left[\frac{x^2}{36}-\frac{E^2}{36}\right]\psi(x) =0,
\label{Dust2}
\end{equation}
Equation (\ref{Dust2}) is formally identical to the
time-independent Schr\"odinger equation for a harmonic oscillator
with unit mass and energy $\lambda$:
\begin{equation}
-\frac{d^{2}\xi}{dx^{2}}+\left[- 2\lambda+w^{2}x^{2}\right]\xi(a)
=0, \label{Dust3}
\end{equation}
where $2\lambda = E^{2}/36$ and $w=1/6$. Therefore, the allowed
values of $\lambda$ are $(n+1/2)w$ and the possible values of $E$
are
\begin{equation}
E_{n}=\pm\sqrt{6(2n+1)}\,\, , \mbox{\hspace{0.8cm}}
n=0,1,2,...\quad . \label{Dust4}
\end{equation}
Thus the stationary solutions are
\begin{equation}
{\Psi}_{n}(a,t)=e^{-iE_{n}t}{\varphi}_{n}\left(12a - E_{n}\right),
\label{Dust6}
\end{equation}
where
\begin{equation}
{\varphi}_{n}(x)=H_n\bigg(\frac{x}{\sqrt{12}}\bigg)e^{-x^2/24}\,\, ,
\label{dust7}
\end{equation}
and $H_n$ are the $n$-th Hermite polynomial.

\begin{table}

 \centering
\begin{tabular}{|c|c|c|c|}
  % after \\: \hline or \cline{col1-col2} \cline{col3-col4} ...
  \hline
   & $k=1$ & $k=0$ & $k=-1$ \\

   \hline
  $E_{0}$&4.660967538\ &\
3.354101966&1.955113416\\

\hline $E_{1}$&11.92641527\ &\ 10.06230590&8.159825054\\

\hline $E_{2}$&18.98089410\ &\ 16.77050983&14.53079652\\

\hline $E_{3}$&25.95270272\ &\ 23.47871376&20.97932418\\

\hline $E_{4}$&32.87827896\ &\ 30.18691770&27.47251428\\

 \hline $E_{5}$&39.77369328\ &\ 36.89512163&33.99515040\\

 \hline $E_{6}$&46.64761142\ &\ 43.60332556&40.53888200\\

 \hline $E_{7}$&53.50529876\ &\ 50.31152949&47.09859109\\

\hline $E_{8}$&60.35022067\ &\ 57.01973343&53.67089213\\

 \hline $E_{9}$&67.18479286\ &\ 63.72793736&60.25341729\\

 \hline $E_{10}$&74.01077406\ &\ 70.43614129&66.84443864\\

\hline $E_{11}$&80.82948903\ &\ 77.14434522&73.44265237\\

 \hline $E_{12}$&87.64196335\ &\ 83.85254916&80.04704775\\

\hline $E_{13}$&94.44900906\ &\ 90.56075309&86.65682363\\

\hline $E_{14}$&101.2512814\ &\ 97.26895702&93.27133299\\

\hline $E_{15}$&108.0493176\ &\ 103.9771610&99.89004490\\

\hline $E_{16}$&114.8435643\ &\ 110.6853649&106.5125177\\

 \hline $E_{17}$&121.6343972\ &\ 117.3935688&113.1383797\\

 \hline $E_{18}$&128.4221359\ &\ 124.1017728&119.7673145\\

 \hline $E_{19}$&135.2070546\ &\ 130.8099767&126.3990507\\

\hline $E_{20}$&141.9893905\ &\ 137.5181806&133.0333530\\

 \hline
\end{tabular}

\caption{The lowest calculated energy levels for the cases $k=0$,
$k=1$, and $k=-1$ in dust dominated Universe (in all cases,
$\Lambda=-15$). As mentioned in the text for every positive
eigenvalue there exist a corresponding negative one with identical
absolute value.}\label{TabDust}
\end{table}

\subsection{Cosmic Strings ($\alpha=-1/3$)}
\begin{figure}
\centering
  % Requires \usepackage{graphicx}
  \includegraphics[width=7cm]{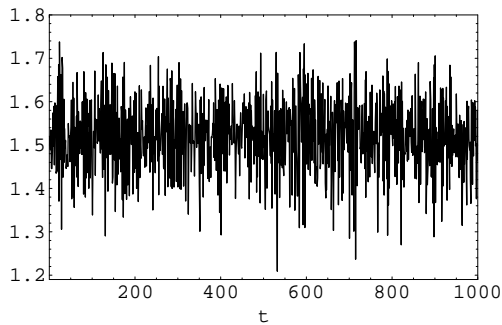}
  \caption{Behavior of the expectation value of the scalar factor for $\Lambda=-15$,
  $k=1$, and $C_n=1,0$ for odd and even $n$,
respectively, in cosmic strings regime.}\label{figcosmic1}
\end{figure}

\begin{figure}
\centering
  % Requires \usepackage{graphicx}
  \includegraphics[width=7cm]{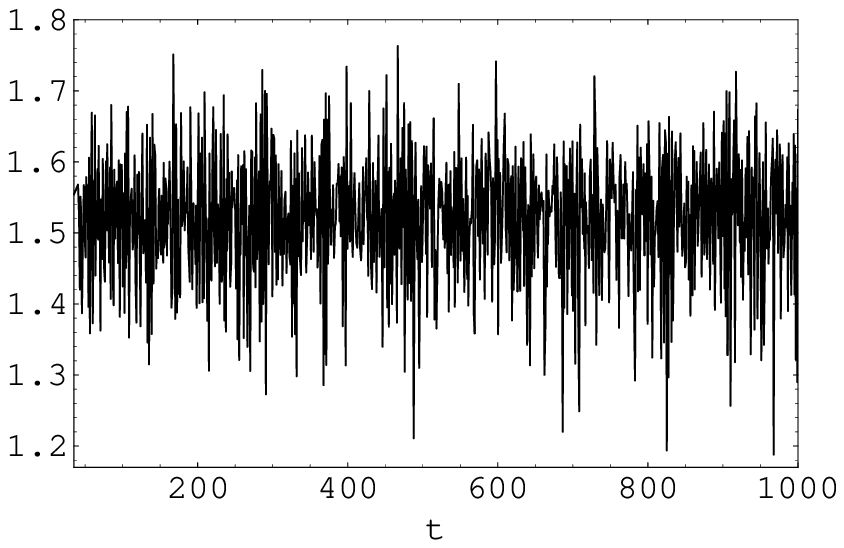}\\
  \caption{Behavior of the expectation value of the scalar factor for $\Lambda=-15$,
  $k=0$, and $C_n=1,0$ for odd and even $n$,
respectively, in cosmic strings regime.}\label{figcosmic0}
\end{figure}

\begin{figure}
\centering
  % Requires \usepackage{graphicx}
  \includegraphics[width=7cm]{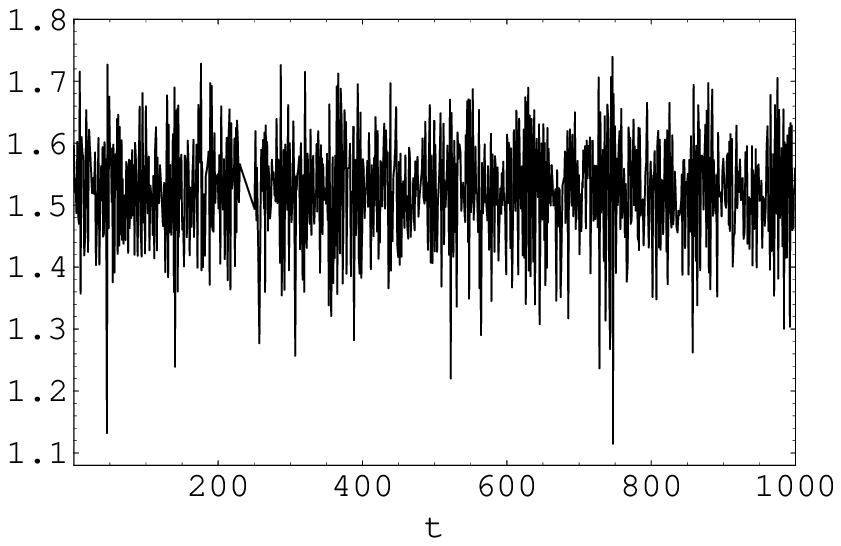}\\
  \caption{Behavior of the expectation value of the scalar factor for $\Lambda=-15$,
  $k=-1$, and $C_n=1,0$ for odd and even $n$,
respectively, in cosmic strings regime.}\label{figcosmic-1}
\end{figure}

In Cosmic Strings dominated Universe time-independent WD equation
has the following form,
\begin{equation}
\label{Cosmic} -\frac{d^2\psi(a)}{da^2} +( 36 ka^2-12\Lambda
a^4)\psi(a) =12E\,a^2\psi(a),
\end{equation}
This differential equation is covariant under parity operator and hence its eigenstates can be
separated into even and odd ones. We can find the energy eigenvalues and eigenvectors of this
equation with ease using SM where $\hat{f}=36 kx^2-12\Lambda x^4 $ and $\hat{g}=12x^2 $ in
comparison with Eq. (\ref{RSM1}). Table \ref{TabCosmicStrings} shows the first 20 eigenvalues for $k=1,0,-1$
respectively. By choosing the first condition of the equation (\ref{boundary}), the resulting wave
packets should consist of only the odd eigenfunctions. Therefore, the coefficients $C_n\,$s in
equation (\ref{wavepacket}) are arbitrary for the odd eigenfunctions zero for the even ones. To be
able to extend the results of Refs. \cite{18,19} for the radiation case to the present one, we
choose the same initial state as their's. That is the odd ones are all chosen to be equal to one.
Figures (\ref{figcosmic1},\ref{figcosmic0},\ref{figcosmic-1}) show the resulting expectation values
of the scale factor $a$, versus $t$ for the various values of $k$.

\begin{table}

 \centering
\begin{tabular}{|c|c|c|c|}
  % after \\: \hline or \cline{col1-col2} \cline{col3-col4} ...
  \hline
   & $k=1$ & $k=0$ &$k=-1$ \\

   \hline
$E_{1}$&11.63722935\ &\ 8.637229353&5.637229353\\

 \hline $E_{2}$&19.36451595\
&\ 16.36451595&13.36451595\\

\hline $E_{3}$&25.54276474\ &\ 22.54276474&19.54276474\\

 \hline $E_{4}$&30.96152618\
&\ 27.96152618&24.96152618\\

\hline $E_{5}$&35.89633715\ &\ 32.89633715&29.89633715\\

 \hline $E_{6}$&40.48441592\
&\ 37.48441592&34.48441592\\

 \hline
$E_{7}$&44.80660320\ &\ 41.80660320&38.80660320\\

 \hline $E_{8}$&48.91557892\
&\ 45.91557892&42.91557892\\

 \hline
$E_{9}$&52.84808309\ &\ 49.84808309&46.84808309\\

\hline $E_{10}$&56.63102596\ &\ 53.63102596&50.63102596\\

\hline $E_{11}$&60.28486358\ &\ 57.28486358&54.28486358\\

\hline $E_{12}$&63.82560554\ &\ 60.82560554&57.82560554\\

 \hline
$E_{13}$&67.26607937\ &\ 64.26607937&61.26607937\\

\hline $E_{14}$&70.61676389\ &\ 67.61676389&64.61676389\\

\hline $E_{15}$&73.88635930\ &\ 70.88635930&67.88635930\\

 \hline
$E_{16}$&77.08218981\ &\ 74.08218981&71.08218981\\

\hline $E_{17}$&80.21050097\ &\ 77.21050097&74.21050097\\

\hline $E_{18}$&83.27676241\ &\ 80.27676241&77.27676241\\

\hline $E_{19}$&86.28668819\ &\ 83.28668819&80.28668819\\

 \hline
$E_{20}$&89.25169642\ &\ 86.25169642&83.25169642\\
\hline

\end{tabular}

\caption{The lowest calculated energy levels for the cases $k=0$,
$k=1$, and $k=-1$ in cosmic strings dominated Universe (in all
cases, $\Lambda=-15$).}\label{TabCosmicStrings}
\end{table}

\subsection{Domain Walls ($\alpha=-2/3$)}
\begin{figure}
\centering
  % Requires \usepackage{graphicx}
  \includegraphics[width=7cm]{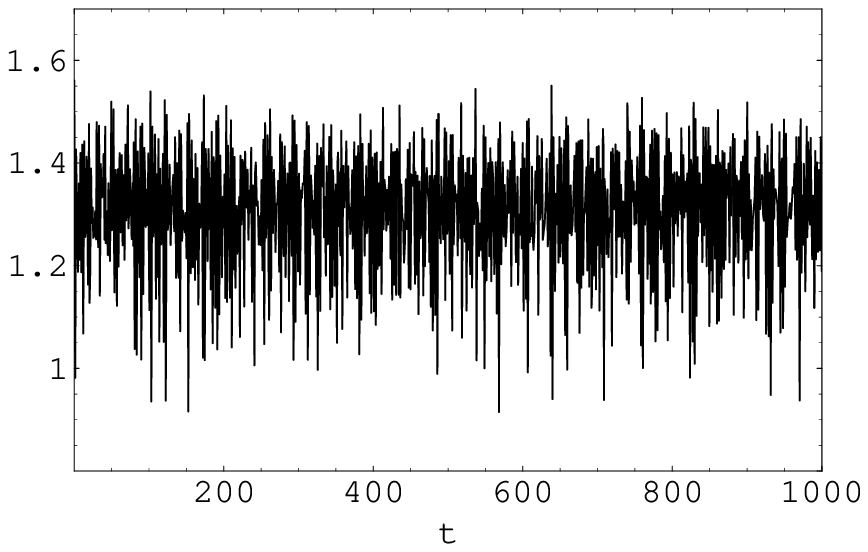}
  \caption{Behavior of the expectation value of the scalar factor for $\Lambda=-15$,
  $k=1$, and $\Psi(a,0) =
\exp(-8(a-1.5)^2)$ in domain walls regime.}\label{figwall1}
\end{figure}

\begin{figure}
\centering
  % Requires \usepackage{graphicx}
  \includegraphics[width=7cm]{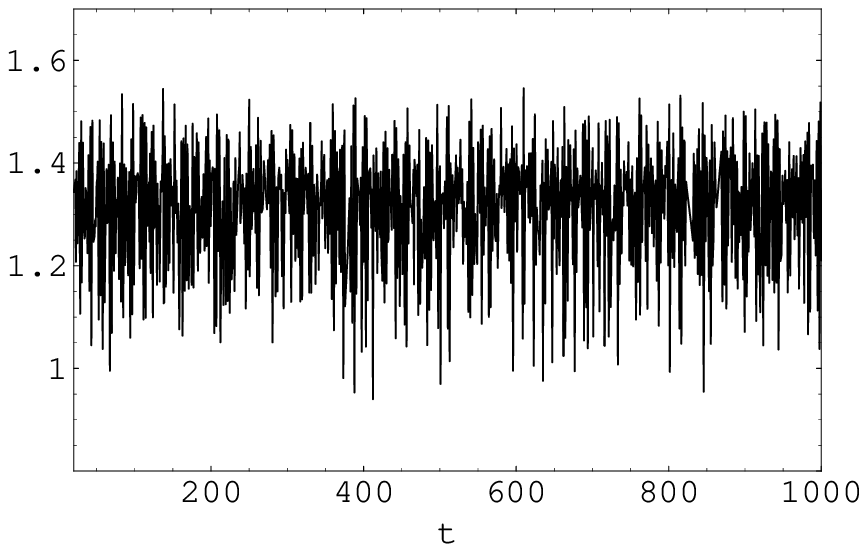}\\
  \caption{Behavior of the expectation value of the scalar factor for $\Lambda=-15$,
  $k=0$, and $\Psi(a,0) =
\exp(-8(a-1.5)^2)$ in domain walls regime.}\label{figwall0}
\end{figure}

\begin{figure}
\centering
  % Requires \usepackage{graphicx}
  \includegraphics[width=7cm]{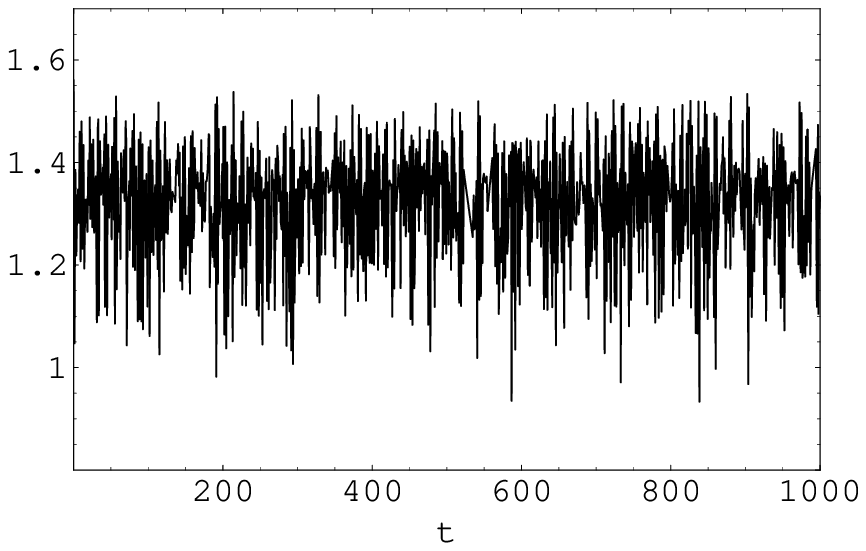}\\
  \caption{Behavior of the expectation value of the scalar factor for $\Lambda=-15$,
  $k=-1$, and $\Psi(a,0) =
\exp(-8(a-1.5)^2)$ in domain walls regime.}\label{figwall-1}
\end{figure}

In Domain Walls dominated Universe ($\alpha=-2/3$) the
time-independent WD equation has the following form,
\begin{equation}
\label{Alpha2/3} -\frac{d^2\psi(a)}{da^2} +( 36 ka^2-12\Lambda
a^4)\psi(a) =12E\,a^3\psi(a),
\end{equation}
We can find the energy eigenvalues and eigenvectors of this equation with ease using SM where
$\hat{f}=36 kx^2-12\Lambda x^4 $ and $\hat{g}=12x^3 $ in comparison with Eq. (\ref{RSM1}). Table \ref{TabAlpha2/3}
shows the first 20 eigenvalues for $k=1,0,-1$ respectively. Note that, for any positive eigenvalues
($E_n^+$), there is an negative counterpart ($E_n^-$) which $E_n^-=-E_n^+$. This case is similar to
the Dust case and in particular its differential equation is not covariant under Parity Operator
and therefore, its eigenfunctions can not in general satisfy either of the conditions stated in
equation (\ref{boundary}. However, we can construct wave packets, from linear combinations of the
eigenfunctions, which vanishes at $a=0$ and $t=0$. Then we need to check that the constraints (Eq.
\ref{boundary} remain valid for all $t$ for our choice of initial condition. For example we can
choose the coefficients $C_n\, $s so as to construct a gaussian initial wave packet ($\Psi(a,0)$)
which is centered \textit{e.g.} at $a=1.5$. Figures (\ref{figwall1},\ref{figwall0},\ref{figwall-1})
show the resulting expectation values of the scale factor $a$, versus $t$ for the various values of
$k$. As can seen from the figures these wave packets always satisfy the first boundary condition
(Eq. \ref{boundary}).


\begin{table}

 \centering
\begin{tabular}{|c|c|c|c|}
  % after \\: \hline or \cline{col1-col2} \cline{col3-col4} ...
  \hline
   & $k=1$ & $k=0$ & $k=-1$

   \\\hline
$E_{1}$&17.07778092\ &\ 12.54649750&7.458961879\\

\hline $E_{2}$&21.41791925\ &\ 18.18718412&14.81583000\\

 \hline
$E_{3}$&24.32147857\ &\ 21.57780684&18.75859852\\

\hline $E_{4}$&26.60465935\ &\ 24.14422756&21.63352524\\

\hline $E_{5}$&28.52359398\ &\ 26.25664404&23.95258205\\

\hline $E_{6}$&30.19749314\ &\ 28.07460050&25.92257362\\

 \hline $E_{7}$&31.69296487\
&\ 29.68328712&27.64978790\\

 \hline
$E_{8}$&33.05153398\ &\ 31.13420271&29.19682156\\

 \hline $E_{9}$&34.30107228\
&\ 32.46114131&30.60396527\\

\hline $E_{10}$&35.46131513\ &\ 33.68761746&31.89883581\\

\hline $E_{11}$&36.54683096\ &\ 34.83073057&33.10127373\\

 \hline
$E_{12}$&37.56911074\ &\ 35.90336205&34.22606529\\

 \hline $E_{13}$&38.54199370\
&\ 36.91577767&35.28463557\\

 \hline
$E_{14}$&39.50080454\ &\ 37.87984925&36.28735433\\

\hline $E_{15}$&40.512127\ &\
38.825301&37.253593\\

\hline $E_{16}$&41.627809\ &\
39.812705&38.232631\\

\hline $E_{17}$&42.859939\ &\
40.895275&39.288909\\

\hline $E_{18}$&44.203790\ &\
42.088418&40.454402\\

\hline $E_{19}$&45.653137\ &\
43.388462&41.730668\\

\hline $E_{20}$&47.203213\ &\
44.789203&43.111449\\

\hline

\end{tabular}
\caption{The lowest calculated energy levels for the cases $k=0$,
$k=1$, and $k=-1$ in domain walls dominated Universe (in all
cases, $\Lambda=-15$). As mentioned in the text for every positive
eigenvalue there exist a corresponding negative one with identical
absolute value.}\label{TabAlpha2/3}
\end{table}

It is important to note that we have repeated the simulations for
all cases ($\alpha=1/3,0,-1/3,-2/3$) with other values of $\Lambda$
and different initial conditions (subject to $\Psi(0,0)=0$). In
particular, we have also repeated simulations for
$\Lambda={-10,-12.5,-17.5,-20}$ rather than $\Lambda=-15$ which
studied in detail, and found the corresponding eigenvalues and
eigenfunctions with desired accuracy. Moreover, we chose other
initial conditions in the form
$\Psi(a,0)=\exp(-\gamma(a-a_0)^{\delta})$ with various choices of
$\gamma$ ($2,5,10,20$), $\delta$ ($2,4,6$), and $a_0$
($1,1.2,1.4,1.6$). We found that, for all these cases the behavior
of the expectation value of the scale factor is similar to ones
depicted in Figs. \ref{figRad1} to \ref{figwall-1} and never tends
to the singular point.

\section{Conclusions}\label{sec5}
In this work we have investigated closed, flat, and open
minisuperspace FRW quantum cosmological models ($k=1,0,-1$) with
perfect fluid for the radiation, dust, cosmic strings, and domain
walls dominated Universes ($\{\alpha=1/3,0,-1/3,-2/3\}$,
respectively). The use of Schutz's formalism for perfect fluids
allowed us to obtain a Schr\"odinger-like WD equation in which the
only remaining matter degree of freedom plays the role of time. We
have used Spectral Method and obtained accurate results for the
eigenfunctions and eigenvalues. Physically acceptable wave packets
were constructed by appropriate linear combination of these
eigenfunctions. The time evolution of the expectation value of the
scale factor has been determined in the spirit of the many-worlds
interpretation of quantum cosmology. Since the expectation values of
the scale factors for the cases considered here never tend to the
singular point, we have an initial indication that these models may
not have singularities at the quantum level. The similar conclusions
have been obtained on general grounds in \cite{14} and for the
radiation case in \cite{18}.

\section*{Acknowledgements}
Authors thank H.~R.~Sepangi for useful discussions and comments.
\pagebreak


\newpage

\begin{thebibliography}{100}
\bibitem{1} B.~S.~DeWitt, Phys.~Rev.~{\bf160}, 1113 (1967).
\bibitem{2} V.~N.~Melnikov and V.~A.~Reshetov, {\it in:\/} Abstr.~8 All-Union Conference on Elementary Particles (Uzhgorod).~Kiev, ITP, 117 (1971).
\bibitem{3} Yu.~N.~Barabanenkov and V.~A.~Pilipenko, {\it in:\/} Abstr.~8 All-Union Conference on Elementary Particles (Uzhgorod).~Kiev, ITP, 117 (1971).
\bibitem{4} C.~W.~Misner, Phys.~Rev.~{\bf 186}, 1419 (1969).
\bibitem{5} Yu.~N.~Barabanenkov, {\it in:\/} Abstr.~3 Soviet Gravitational Conference, Yerevan (1972).
\bibitem{6} M.~I.~Kalinin, V.~N.~Melnikov, Grav.~Cosmol.~\textbf{9}, 227 (2003).
\bibitem{7} R.~Arnowitt, S.~Deser and C.~W.~Misner, {\it Gravitation: An Introduction to Current Research}, edited by L.~Witten, Wiley, New York (1962).
\bibitem{8} C.~J.~Isham, {\it Canonical quantum gravity and the problem of time}, arXiv:gr-qc/9210011.
\bibitem{9} F.~J.~Tipler, Phys.~Rep.~{\bf137}, 231 (1986).
\bibitem{10} C.~Kiefer, Phys.~Rev.~D {\bf38}, 1761 (1988).
\bibitem{11} B.~F.~Schutz, Phys.~Rev.~D {\bf2}, 2762 (1970).
\bibitem{12} B.~F.~Schutz, Phys.~Rev.~D {\bf4}, 3559 (1971).
\bibitem{pedramPLB} P.~Pedram, S.~Jalalzadeh and S.~S.~Gousheh, Phys.~Lett.~B \textbf{655}, 91 (2007), arXiv:0708.4143.
\bibitem{pedramCQG2} P.~Pedram, S.~Jalalzadeh and S.~S.~Gousheh, Class.~Quantum Grav.~\textbf{24}, 5515 (2007), arXiv:0709.1620.
\bibitem{pedramIJTP} P.~Pedram, S.~Jalalzadeh and S.~S.~Gousheh, Int.~J.~Theor.~Phys.~doi:10.1007/s10773-007-9436-9, arXiv:0705.3587.
\bibitem{pedramPLB2} P.~Pedram and S.~Jalalzadeh, Phys.~Lett.~B (2007), doi: 10.1016/j.physletb.2007.11.013, arXiv:0705.3587.
\bibitem{13} M.~J.~Gotay and J.~Demaret, Phys.~Rev.~D {\bf 28}, 2402 (1983).
\bibitem{14} V.~G.~Lapchinskii and V.~A.~Rubakov, Theor.~Math.~Phys.~{\bf 33}, 1076 (1977).
\bibitem{15} F.~G.~Alvarenga and N.~A.~Lemos, Gen.~Rel.~Grav.~{\bf 30}, 681 (1998).
\bibitem{16} J.~Acacio de Barros, N.~Pinto-Neto and M.~A.~Sagioro-Leal, Phys.~Let.~A {\bf 241}, 229 (1998).
\bibitem{17} F.~G.~Alvarenga, J.~C.~Fabris, N.~A.~Lemos, G.~A.~Monerat, Gen.~Rel.~Grav.~{\bf 34}, 651 (2002).
\bibitem{18} G.~A.~Monerat, E.~V.~C.~Silva, G.~Oliveira-Neto, L.~G.~F.~Filho, and N.~A.~Lemos, Phys.~Rev.~D \textbf{73}, 044022 (2006).
\bibitem{19} P.~Amore, A.~Aranda, M.~Cervantes, J.~L.~D\`{\i}az-Cruz, Francisco M.~Fern\`{a}ndez, Phys.~Rev.~D \textbf{75}, 068503 (2007), arXiv:gr-qc/0611029.
\bibitem{boyd} J.~P.~Boyd, {\it Chebyshev and Fourier Spectral Methods},  second ed., Dover, New York, (2001).
\bibitem{SMP}P.~Pedram, M.~Mirzaei, S.~S.~Gousheh, arXiv:math-ph/0611008.
\bibitem{25} P.~Pedram, M.~Mirzaei, S.S.~Gousheh, Comput.~Phys.~Commun.~\textbf{176}, 581 (2007).
\bibitem{assad} M.~J.~Assad and J.~A.~S.~Lima, Gen.~Rel.~Grav.~{\bf 20}, 527 (1988).
\bibitem{nivaldo} N.~A.~Lemos, J.~Math.~Phys.~{\bf 37}, 1449 (1996).
\bibitem{22} H.~Everett, III, Rev.~Mod.~Phys.~{\bf 29}, 454 (1957).
\bibitem{Chhajlany1} S.~C.~Chhajlany and V.~N.~Malnev, Phys.~Rev.~A \textbf{42}, 3111 (1990).
\bibitem{Chhajlany2} S.~C.~Chhajlany, D.~A.~Letov, and V.~N.~Malnev, J.~Phys.~A \textbf{24}, 2731 (1991).
\bibitem{Amore} P.~Amore, J.~Phys.~A \textbf{39}, L349 (2006).
\bibitem{Stevenson} P.~M.~Stevenson, Phys.~Rev.~D \textbf{23}, 2916 (1981).
\bibitem{Reply} N.~A.~Lemos, G.~A.~Monerat, E.~V.~C.~Silva, G.~Oliveira-Neto, and L.~G.~Ferreira Filho, Phys.~Rev.~D \textbf{75}, 068504 (2007).
\end{thebibliography}
\end{document}